\documentclass[12pt,preprint]{aastex}
\def\kms{km~s$^{-1}$}

\slugcomment{{\it ISRN Astronomy and Astrophysics}, vol.\ 2011, id \# 958973 \\ 
doi: 10.5402/2011/958973}
\shorttitle{Locations of Satellite Galaxies}
\shortauthors{\'Ag\'ustsson and Brainerd}

\begin{document}

\title{Locations of Satellite Galaxies in the Two Degree Field Galaxy
Redshift Survey}

\author{Ing\'olfur \'Ag\'ustsson and Tereasa G.\ Brainerd}
\affil{Boston University, Department of Astronomy, 725
Commonwealth Ave., Boston, Massachusetts, USA 02215}
\email{ingolfur@bu.edu, brainerd@bu.edu}

\begin{abstract}
We compute the locations of satellite galaxies
in the Two Degree Field Galaxy Redshift Survey using two
sets of selection criteria and three sources of photometric data.
Using the SuperCOSMOS $r_F$ photometry, we find that
the satellites are located preferentially
near the major axes of their hosts, and the anisotropy
is detected at a highly-significant level (confidence levels of
99.6\% to 99.9\%).
The locations
of satellites that have high velocities relative to their hosts
are statistically indistinguishable from the locations of
satellites that have low velocities relative to their hosts.
Additionally, satellites with passive
star formation are distributed anisotropically about their
hosts (99\% confidence level), while the locations of
star-forming satellites are consistent with an isotropic distribution.
These two distributions are, however, statistically indistinguishable.
Therefore it is not correct to interpret this as evidence that
the locations of the star-forming satellites are intrinsically
different from those of the passive satellites.
\end{abstract}


\section{Introduction}


The existence of massive halos of dark matter around large, bright 
galaxies is well-accepted.  However, at present there are 
relatively few direct observational
constraints on the sizes and shapes of these dark matter
halos. The most popular theory for structure formation
in the universe, known as Cold Dark Matter (CDM), predicts that the dark matter halos
extend to radii that are at least an order of magnitude greater than the radii
of the visible galaxies (see, e.g., [1] and references therein).  In
addition, CDM predicts that the dark matter halos of galaxies
are not spherical; instead they
are triaxial in shape (e.g., [2], [3], [4], [5], [6]).  

In principle,
the locations of small, faint satellite galaxies, measured with respect to the 
major axes of the large, bright, ``host'' galaxies that they orbit,
have the potential to provide strong constraints on the dark matter halos
that surround the hosts, as well as on the relationships of
the luminous hosts to their dark matter halos.  Recent studies of
satellite galaxies from modern redshift surveys have shown that,
when their locations are averaged over the entire population, 
the satellites of relatively isolated host galaxies have
a preference for being located near the major axes of their hosts (e.g., [7],
[8], [9], [10], [11]).
The observed locations of the satellite
galaxies in the Sloan Digital Sky Survey (SDSS; [12]) are also
known to depend upon 
various physical properties of the hosts and satellites
(e.g., [9], [11], [13]).  
The satellites of the SDSS host galaxies that have the reddest colors,
highest stellar masses, and lowest specific star formation rates (SSFR) show a pronounced
tendency for being located near the major axes of their hosts.  
(Note: the SSFR has units of yr$^{-1}$ and is defined to be the ratio of the star formation rate 
in the galaxy to its
stellar mass; see [11].) 
On the other hand,
the satellites of the SDSS host galaxies that have the bluest colors, 
lowest stellar masses,
and highest SSFR are distributed isotropically around their
hosts.   The SDSS satellite galaxies that have the reddest colors, highest stellar masses,
and lowest SSFR also show a strong preference for being
located near the major axes of their hosts, while the SDSS satellite
galaxies that have the
bluest colors, lowest stellar masses, and highest SSFR
show little to no anisotropy in their locations.
The alignment of the satellites of relatively isolated SDSS host galaxies is
also known to be similar to the alignment of satellites with the central galaxies 
of relatively isolated SDSS galaxy groups, where
the strongest alignment is found for red central galaxies and their red satellites,
while no significant satellite alignment is detected for groups that have blue central
galaxies (e.g., [14]).

From a theoretical standpoint, one would expect that if the dark matter
halos of large, bright galaxies consist of CDM, then the
locations of the satellite galaxies should reflect
the deviations of the halo potentials from pure spherical symmetry.
Simulations of structure formation in $\Lambda$CDM universes have shown that,
in projection on the sky, the locations of the satellite galaxies
trace the shapes of their hosts' dark matter halos rather well (e.g., [15]).
However (and crucially), from an observational
standpoint, the expected non-spherical distribution of satellite
galaxies will only
manifest in an observational data set if mass and light are reasonably well-aligned
within the hosts.  
In other words, the satellites should trace the dark mass associated with their hosts, 
but not
necessarily the luminous mass associated with their hosts (i.e., since the dark
mass exceeds the luminous mass by $\sim 2$ orders of magnitude).

If the halos of the hosts are triaxial, and if one could simply
use the symmetry axes of the hosts' dark matter halos (as projected on the sky) to
define the geometry of the problem,
one would naturally expect to observe an anisotropy in the locations of satellite
galaxies such that the satellites are found preferentially
close to the major axes of
their hosts' dark matter halos.  If there is a substantial misalignment between the
projected major axes of the luminous host galaxies and their dark matter halos,
however,
one would expect to observe little to no anisotropy in the locations of the
satellites.  Using simple prescriptions for embedding luminous host galaxies
within their dark matter halos, [11] showed that the observed dependences of 
SDSS satellite locations on various host properties can be easily reproduced if 
mass and
light are aligned in the elliptical
hosts (i.e., luminous ellipticals are effectively miniature versions of their 
dark matter halos), while
the disk hosts are instead oriented such that their angular momentum vectors 
are aligned with the
net angular momentum vectors of their halos.  The angular momentum
alignment for the disk hosts and their halos introduces a
significant misalignment of mass and light (e.g., [16]), resulting
in the satellites of disk hosts being distributed much more isotropically than
the satellites of elliptical hosts.

One of the difficulties with observational samples of 
host galaxies and their satellites is the presence of ``interlopers'' (i.e., 
``false'' satellites) in the data.  Since the distances to the galaxies are generally
unknown, hosts and satellites are selected from redshift surveys via a set
of redshift space proximity criteria.  Typically, satellite galaxies must be
located within a projected distance $r_p \le 500$~kpc of their host, and the
line of sight velocity difference between a host and its satellite must
be $| \delta v| \le 500$~km~s$^{-1}$.  From simulations
in which hosts and satellites were selected using criteria
that are identical to the redshift space criteria used for observational data sets,
it is known that the majority of
objects that are selected as satellite galaxies are, in fact, located physically
nearby a host
galaxy.  However, a substantial number of objects that are selected as satellites
are located physically far away
from a host galaxy and are, therefore, interlopers (i.e., not genuine
satellites; see [11]).
When investigating the properties of the satellite
population, the interlopers are a source of noise and ideally one would
eliminate them from the sample if at all possible.  This can be done
in a simulation since the 3-dimensional locations of all of the objects are known, 
but it
is not obvious how or if this
can be accomplished in an observational data set.

So far, the only direct attempt to eliminate interlopers from an observational
study of
the locations of satellite galaxies is the work of [7]. In their study,
[7] computed the 
locations of the satellites of relatively isolated host galaxies in the 
Two Degree Field Galaxy Redshift Survey (2dFGRS; [17], [18]).  
In order to address the detrimental effects of interloper contamination, [7]
argued that if they divided their host-satellite sample by relative
line of sight
velocity, $| \delta v|$,  the set of host-satellite pairs that had the largest
observed values of $| \delta v|$ should suffer greater
interloper contamination than the set of host-satellite pairs that had the
smallest observed values of $| \delta v|$.  
That is, [7] anticipated that the peak
of the observed relative velocity distribution, $P(\delta v)$, would be dominated
by genuine satellites, while the tails of the distribution would be
dominated by interlopers (for which the observed values of $| \delta v|$
would be largely attributable to the Hubble flow).  Therefore, [7]
divided their sample of hosts and satellites into a ``low relative velocity''
sample ($| \delta v | < 160$~km~s$^{-1}$) and a ``high relative velocity''
sample ($| \delta v | > 160$~km~s$^{-1}$), expecting that the low relative
velocity sample would suffer much less interloper contamination
in comparison to the high relative velocity sample.
Within both the full sample and
the high relative velocity sample, [7]
found no evidence for any anisotropy in the locations of the satellite galaxies.
However, in the sample with $| \delta v | < 160$~km~s$^{-1}$, [7] reported
a preference for the satellites to be located near the major axes of 
their hosts (see [10], the erratum to [7]).  Within this low relative
velocity sample, [7] found that the ratio of ``planar'' ($\phi < 30^\circ$)
to ``polar'' ($\phi > 60^\circ$) satellite locations was
$f = N_{< 30}/N_{> 60} = 1.25\pm 0.06$ and that the distribution of satellite
locations was fitted well by a double cosine function with amplitude 
$A = 0.12 \pm 0.04$.

In their analysis, [7] did not directly determine whether the results
for the satellite locations in the low velocity sample were statistically
distinct from the results for the satellite locations in the high velocity
sample.  That is, given the small number statistics with which [7] were
working, it is entirely possible that the distribution of satellite locations
in their high velocity sample was
consistent with being drawn from the same
parent population as the distribution of satellite locations in the low
velocity sample.  Therefore, it is not clear that their result should be
interpreted as evidence that the satellites in the high velocity sample 
are distributed isotropically about their hosts, while the satellites in
the low velocity sample are distributed anisotropically about their hosts.
Rather, all that can be concluded about the high relative velocity sample
in [7]
is that the null hypothesis of a uniform distribution could not be ruled out.
The fact that the null hypothesis could not be ruled out may be due to
the locations of the satellites in the high velocity sample having an
intrinsically isotropic distribution.
On the other hand, it could
also be due to the size of the sample being too small to detect an intrinsic
anisotropy in the presence of a significant amount of
noise (i.e., this is a pair-counting problem
that is dominated by $\sqrt{N}$ statistics).

At the time [7] were doing their work, little was known about
the distribution of the interlopers relative to the host
galaxies and, for the most part, interlopers
were simply assumed to be a population of objects that were selected
at random (see, e.g., [19], [20], [21]).
However, careful analyzes of the interloper population
from simulations has shown that the interlopers are far from being a random
population.  Instead, along the line of sight,
most interlopers are located within a distance
of $\pm 2$~Mpc of a host (i.e., a distance far less than the $\sim 7$~Mpc one
would expect from the Hubble flow, given a maximum host-interloper velocity
difference of $| \delta v | = 500$~\kms; see [22]).  
In addition, the probability distribution
of relative velocities, $P(\delta v)$, for the hosts and interlopers 
reaches a maximum at $\delta v = 0$ (e.g., [22], [23]).  
The distribution of relative velocities for host-interloper pairs is,
in fact, quite similar to the distribution of relative velocities for 
pairs of hosts and their genuine satellites.  Therefore, 
interlopers are almost as likely to
have low velocities relative to the host galaxies as are the genuine satellites.
In retrospect, then, it is not clear that the original velocity cut that [7]
imposed in their analysis is well-motivated, nor that there should be any 
significant difference in the locations of satellites with low velocities 
relative to their hosts 
and the locations of satellites with high velocities relative to their hosts.

Here we revisit the question of the locations of satellite galaxies in
the 2dFGRS.  We first adopt the selection criteria of [7] to obtain
a host-satellite sample, and we compute the satellite locations
using three different sets of photometry 
for the galaxies.  We next adopt the selection criteria that we used
in a previous study of the locations of satellite galaxies in the SDSS
(e.g., [11]), 
and we focus our analysis on the hosts and satellites found using
the SuperCOSMOS scans of the $r_F$ plates. 
In all cases we determine whether the satellite
locations in a low relative velocity sample of the data
can be distinguished from
the satellite locations in a high relative velocity sample.  Finally, using
the sample obtained with the SDSS selection criteria, we investigate
the effect of star formation rate on the observed locations of the 
2dFGRS satellites.
Throughout we adopt cosmological parameters $\Omega_{m0} = 0.25$,
$\Omega_{\Lambda 0} = 0.75$ and 
$H_0 = 73$ km~s$^{-1}$~Mpc$^{-1}$.

\section{Two Degree Field Galaxy Redshift Survey}

The Two Degree Field Galaxy Redshift Survey\footnote{\tt 
http://msowww.anu.edu.au/2dFGRS/} is
a publicly-available redshift survey that covers $\sim 5$\% of the sky.  The
target objects in the survey were selected in the $b_J$ band from the
Automated Plate Measuring (APM) galaxy survey and extensions to the 
survey (see [24] and [25]).  The photometry of the APM galaxy
survey was based on scans of the UK Schmidt Telescope photographic survey
plates obtained in blue ($b_J$) and red ($r_F$) spectral bands.  Although
the APM did not complete the scans of the $r_F$ plates, the SuperCOSMOS measuring
machine was ultimately used to make independent scans of both the $b_J$
and $r_F$ plates (see [26] and [27])\footnote{\tt http://www-wfau.roe.ac.uk/sss/}.
In addition to providing photometry in two spectral bands,  [28]
report that the SuperCOSMOS scans yielded improved linearity and smaller
random errors in comparison to the original APM scans.  The final data
release of the 2dFGRS contains 245,591 galaxies, of which
233,251 have good quality spectra ($Q \ge 3$).  
Here we use the final 2dFGRS data release and, specifically, we
use the data for the best spectrum of each object (i.e., the ASCII catalog
{\bf ``best.observations.idz''}, which contains the 2dFGRS spectral information,
as well as the 
photometric information from the APM scans, for 231,178 sources with
$Q \ge 3$ and extinction-corrected magnitudes $b_J \le 19.45$). 
Additionally, we use the 2dFGRS database to retrieve the
apparent magnitudes, the galaxy shape
parameters (semi-major and semi-minor axes) and the
galaxy position angles for the 
SuperCOSMOS scans of the $b_J$ and $r_F$ plates.

Spectral types for the
galaxies are quantified by the parameter $\eta$, which can be interpreted
as an indicator of the amount of star formation in the galaxy
(e.g., [29]).  Rest-frame colors for the 2dFGRS galaxies can
be obtained by using the relationship
\begin{equation}
(b_J-r_F)_0 = b_J - r_F - K(b_J) + K(r_F), 
\end{equation}
where $K(b_J)$ and $K(r_F)$
are color-dependent K-corrections from [30].

\section{Locations of Satellites: Sample 1}

We begin by obtaining
hosts and satellites from the 2dFGRS using selection
criteria that are identical to the criteria used by [7].  
In selecting this host-satellite sample, we use the photometric
parameters from the APM scans of the $b_J$ plates, as did [7].  
Here host galaxies must
have redshifts $z < 0.1$, absolute magnitudes $B_J < -18$, and image
ellipticities $e = 1 -b/a \le 0.1$. 
In addition, host galaxies must
be relatively isolated within their local regions of space.  In order
for a host to qualify as being relatively isolated, its $B_J$
magnitude must be at least one magnitude brighter than 
any other galaxy that is found within a projected radius of
$r_p < 700$~kpc and line of sight velocity difference $| \delta v | < 
1000$~\kms.  Satellites are galaxies that have absolute
$B_J$ magnitudes that are at least two magnitudes fainter than their
host, are 
found within projected radii $r_p < 500$~kpc and have line of
sight velocity differences $|\delta v | < 500$~\kms~relative to their hosts.
In order to exclude host-satellite
systems that are likely to be groups of galaxies,
we reject all host-satellite systems that meet the above criteria, 
but which contain 5
or more satellites (see also [7]).  
After all of the above restrictions are imposed,
our first sample consists of 1,725 hosts and 2,594 satellites.  The size
of our sample is slightly larger than that of [7] (who had 1,498
hosts), and the difference in sample size is likely
attributable to small differences in the implementation of the 
selection criteria.

We define the location of a satellite, $\phi$, to be the angle 
between the major axis of its host galaxy and the direction vector
on the sky that connects the centroids of the host and satellite.
Since we are only interested in determining whether the satellites
are found preferentially close to either the major or minor axes
of their hosts, we restrict $\phi$ to the range $0^\circ \le \phi
\le 90^\circ$.  Therefore, ``planar alignment'' corresponds to a
mean satellite location $\left< \phi \right> < 45^\circ$ and ``polar
alignment'' corresponds to a mean satellite location 
$\left< \phi \right> > 45^\circ$.

Shown in the top panels of Figure~1 are the differential 
and cumulative
probability distributions for the 
satellite locations in our first sample (panels a) and b), respectively). 
Here the centroids of the hosts and satellites,
as well as the position angles of the host
galaxies, are taken from the APM scans of the $b_J$ plates.  Error bars for
$P(\phi)$ were computed using 1,000 bootstrap
resamplings of the data. Also shown in the top
panels of Figure~1 are the mean satellite location,
the median satellite location, the confidence level at which the $\chi^2$
test rejects a uniform distribution for $P(\phi)$, 
and the confidence level at which the 
Kolmogorov-Smirnov (KS) test rejects a uniform distribution for $P(\phi \le
\phi_{\rm max})$.  From Figure~1a) and 1b), then, the satellite locations
in our first sample
are consistent with an isotropic distribution.  Following [7] we
also compute the planar-to-polar ratio, $f = N_{< 30} / N_{> 60} = 1.08 \pm
0.05$, which again is consistent with an isotropic distribution. 

\begin{figure}
\centerline{\scalebox{0.90}{\includegraphics{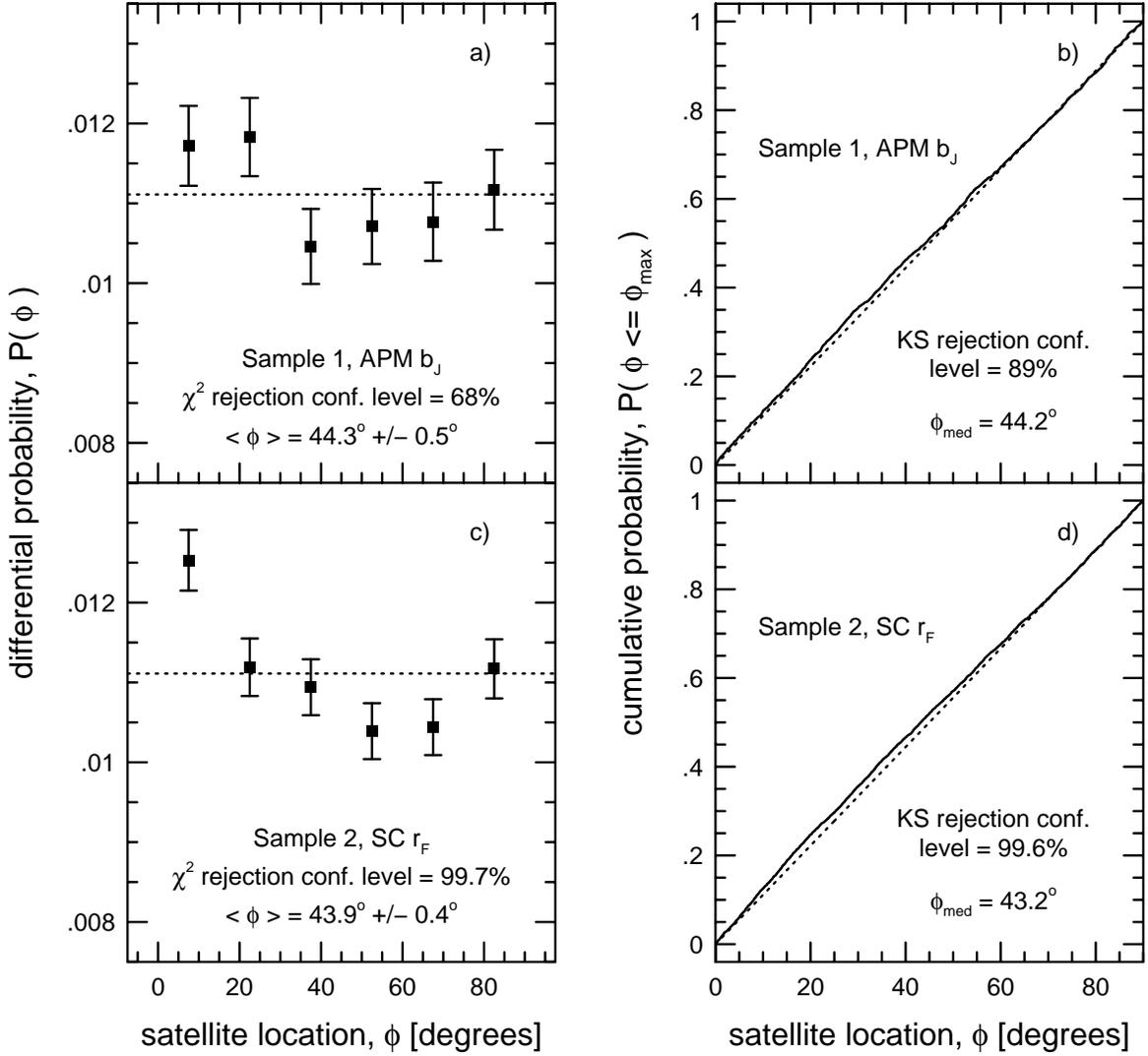} } }%
\vskip -0.0cm%
\caption{Probability distributions for the
locations of satellite galaxies in the 2dFGRS.
{\it Top:} Results for our first sample, where 
the APM scans of the $b_J$ plates, the selection
criteria from [7], and all host-satellite pairs
are used in the calculations.  
{\it Bottom:} Results for our second sample, where
the SuperCOSMOS scans of the $r_F$ plates, the
selection criteria from [11], and all host-satellite
pairs are used in the calculations.  {\it Left:} Observed
differential probability
distribution (data points with error bars). 
Dotted lines show the expectation for a
uniform (i.e., isotropic) distribution.  Also shown are the mean satellite location
and the confidence level at which the $\chi^2$ test rejects the 
uniform distribution.  {\it Right:} Observed cumulative probability distribution
(solid lines) and the expectation for a uniform distribution (dotted lines). Also
shown are the median satellite location and the confidence level 
at which the KS test rejects the uniform distribution.
}
\label{fig1}
\end{figure}

Next, and again following [7],
we divide our first sample into a ``low relative velocity'' sample
($| \delta v | < 160$~\kms; 1,209 hosts and 1,514 satellites) and a
``high relative velocity'' sample ($| \delta v | > 160$~\kms; 855
hosts and 1,080 satellites), and we repeat the analysis above.  Unlike [7],
however, we find no statistically significant
indication that the satellites in either velocity
sample are distributed anisotropically around their hosts.  
Further, a two-sample KS test that compares $P(\phi \le \phi_{\rm max})$
for the low relative velocity sample to $P(\phi \le \phi_{\rm max})$ for
the high relative velocity sample finds that the two distributions
are statistically identical.  That is, the two-sample KS test rejects the 
null hypothesis that
the two distributions are drawn from the same parent distribution
at a confidence level of 18\%.
We summarize our
results in lines 1--3 of Table~1, where $P_{KS}$ is the confidence level at which the
KS test rejects a uniform distribution for the satellite locations,
$\left< \phi \right>$ is the mean satellite location, 
$\phi_{\rm med}$
is the median satellite location, and $f$ is the planar-to-polar
ratio.  The error bound on $\left< \phi \right>$ is 
the standard deviation in the mean, and the error bounds on $\phi_{\rm med}$
and $f$ are 68\%
confidence bounds obtained from 2,000 bootstrap resamplings of the data.

Lastly, we repeat our analysis using the same hosts and satellites
as above, but we now obtain the 
host galaxy position angles 
from the SuperCOSMOS scans of the $b_J$ and $r_F$ plates.  Using the 
host position angles from the
SuperCOSMOS scans has no affect on our conclusions above; in all
cases the locations of the satellites are consistent with an
isotropic distribution.
We summarize our 
results for the locations of the satellites from the 
SuperCOSMOS scans in Table~1, lines 4--6 ($b_J$) and lines 7--9 ($r_F$).  
A two-sample KS test that compares $P(\phi \le \phi_{\rm max})$
for the low relative velocity sample to $P(\phi \le \phi_{\rm max})$ for
the high relative velocity sample finds that, for the $b_J$
SuperCOSMOS scans, the two distributions
are statistically indistinguishable (KS rejection confidence level of 63\%).
Similarly, 
a two-sample KS test that compares $P(\phi \le \phi_{\rm max})$
for the low relative velocity sample to $P(\phi \le \phi_{\rm max})$ for
the high relative velocity sample finds that, for the $r_F$
SuperCOSMOS scans, the two distributions
are also statistically indistinguishable (KS rejection confidence level of 7\%).

\bigskip
\bigskip
\centerline{\bf Table 1: Satellite Locations in the 2dFGRS}
\bigskip
\centerline{
\begin{tabular}{ccccccc}
\hline\hline
Sample & scan &
$| \delta v|$  & $P_{KS}$ & $\left< \phi \right>$ (degrees) & $\phi_{\rm med}$ (degrees) & $f$ \\ \hline
1 & APM $b_J$ & $< 500$~\kms & 89\% & $44.3 \pm 0.5$ & $44.2^{+0.7}_{-1.1}$ & $1.08 \pm 0.05$ \\
1 & APM $b_J$ & $< 160$~\kms & 91\% & $44.0 \pm 0.7$ & $43.9^{+0.9}_{-1.6}$ & $1.11 \pm 0.07$ \\
1 & APM $b_J$ & $> 160$~\kms & 9\% & $44.7 \pm 0.8$ & $44.7^{+1.2}_{-1.8}$ & $1.04 \pm 0.08$ \\
\hline
1 & SC $b_J$ & $< 500$~\kms & 86\% & $44.3 \pm 0.5$ & $44.0^{+1.0}_{-0.9}$ & $1.08 \pm 0.05$ \\
1 & SC $b_J$ & $< 160$~\kms & 90\% & $43.8 \pm 0.7$ & $43.3^{+0.9}_{-1.4}$ & $1.13 \pm 0.07$ \\
1 & SC $b_J$ & $> 160$~\kms & 6\% & $44.9 \pm 0.8$ & $45.3^{+2.0}_{-1.3}$ & $1.00 \pm 0.08$ \\
\hline
1 & SC $r_F$ & $< 500$~\kms & 90\% & $44.4 \pm 0.5$ & $44.1^{+0.9}_{-1.0}$ & $1.10 \pm 0.05$ \\
1 & SC $r_F$ & $< 160$~\kms & 83\% & $44.4 \pm 0.7$ & $43.8^{+1.3}_{-1.0}$ & $1.10 \pm 0.07$ \\
1 & SC $r_F$ & $> 160$~\kms & 31\% & $44.5 \pm 0.8$ & $44.3^{+1.6}_{-1.3}$ & $1.08 \pm 0.08$ \\
\hline
2 & SC $r_F$ & $< 500$~\kms & 99.6 \% & $43.9 \pm 0.4$ & $43.2^{+0.7}_{-0.5}$ & $1.10 \pm 0.04$ \\
2 & SC $r_F$ & $< 160$~\kms & 99.2\% & $43.7 \pm 0.5$ & $42.7^{+0.9}_{-1.0}$ & $1.12 \pm 0.05$ \\
2 & SC $r_F$ & $> 160$~\kms & 87\% & $44.3 \pm 0.6$ & $43.6^{+1.0}_{-0.7}$ & $1.07 \pm 0.06$ \\
\hline
\end{tabular}
}

\bigskip
\bigskip
Therefore, at least in case of the selection criteria
adopted by [7], our analysis finds that there is no statistically significant evidence that
the 2dFGRS satellites are distributed anisotropically
around their hosts.
Further, we find that there is no statistically significant evidence that dividing
the host-satellite sample by relative velocity (i.e., low vs. high) 
results in different conclusions about the locations of the satellites.

\section{Locations of Satellites: Sample 2}

In order to compare most directly with our previous work using SDSS galaxies,
we next obtain a host-satellite sample from the 2dFGRS using the selection
criteria from [11].  
Since the SDSS results are based upon $r$-band imaging, and also because the
shapes of galaxies
are generally smoother at longer wavelengths than they are at
shorter wavelengths (i.e.,
the position angles of the host galaxies may be more accurate when
measured at longer wavelengths),   
here we restrict our analysis to the SuperCOSMOS scans of the $r_F$ plates.
The selection criteria that we adopt are 
similar to the selection criteria that we
used to obtain our first sample, but here they are somewhat
more relaxed.
Host galaxies must have $r_F$ magnitudes that are at least
one magnitude brighter than any other galaxy that is found within 
a projected radius $r_p \le 700$~kpc and a line of sight velocity difference
$| \delta v| \le 1,000$~\kms.  Satellite galaxies are objects that, relative
to their hosts, are found within projected radii $r_p \le 500$~kpc, have
line of sight velocity differences $| \delta v | \le 500$~\kms~and have
$r_F$ magnitudes that are at least two magnitudes fainter than their
host.  In addition, the luminosity of each host must
exceed the sum total of the luminosities of its satellites, each host
may have at most 9 satellites, and hosts are restricted to the 
redshift range $0.01 \le z \le 0.15$.  We place no restrictions either on the 
ellipticities of the hosts' images or on their absolute magnitudes.
However, we do require that the hosts and satellites have
good quality spectra ($Q \ge 3$), and that the hosts have well-defined
spectral parameters ($\eta \ne -99.9$).  The latter constraint helps to 
insure that the host galaxies have fairly regular shapes.
This results in 2,947 host galaxies and
4,730 satellites in our second sample (i.e., $\sim 80$\% larger than
our first sample above). 

We assign
rest-frame colors to the 2dFGRS hosts and satellites using Equation (1) above.
Following [30] we define red galaxies to be those with rest-frame
colors $(b_J - r_F)_0 \ge 1.07$ and blue galaxies to be those with
rest-frame colors $(b_J - r_F)_0 < 1.07$.  Following [29] we 
use the value of $\eta$ as a measure of the star formation rate within a
galaxy, from which we define
galaxies with $\eta > -1.4$ to be ``star-forming'' and galaxies with
$\eta \le -1.4$ to be ``passive''.  Although rest-frame color and star formation
rate are strongly correlated (i.e., red galaxies tend to have low star formation
rates, while blue galaxies tend to have high star formation rates) these two 
parameters are not identical.  Within our sample, 11\% of the 
``passive'' hosts have blue rest-frame colors and 28\% of the ``star-forming''
hosts have red rest-frame colors.  Of the 4,332 satellites that
have well-defined spectral parameters, 13\% of the ``passive''
satellites have blue rest-frame colors and 10\% of the ``star-forming''
satellites have red rest-frame colors.

Figure~2 summarizes the basic statistical properties of the 
SuperCOSMOS $r_F$
host-satellite sample obtained using the selection criteria of [11].  The
different panels of Figure~2 show
probability distributions for: 
a) the number of satellites per host,
b) the redshifts of the hosts, c) the $r_F$ apparent magnitudes
of the hosts and satellites, d) the $r_F$ absolute magnitudes
of the hosts and satellites, e) the rest-frame colors of
the hosts and satellites, and f) the spectral types of the hosts and satellites.
From Figure~2, then, our host sample is dominated by red, passive galaxies
while our satellite sample is dominated by blue, star-forming galaxies.
This is in good agreement with our previous results for the SDSS (e.g., [11]).
For comparison, Figure~3 shows the basic statistical properties for the
hosts and satellites of Sample~1 (see Section~3 above).  Aside from differences
that are due to different imposed cutoffs (i.e., maximum number of satellites
and host galaxy redshift range), the statistical
properties of the hosts and satellites are very similar for our two samples.

\begin{figure}
\centerline{\scalebox{0.90}{\includegraphics{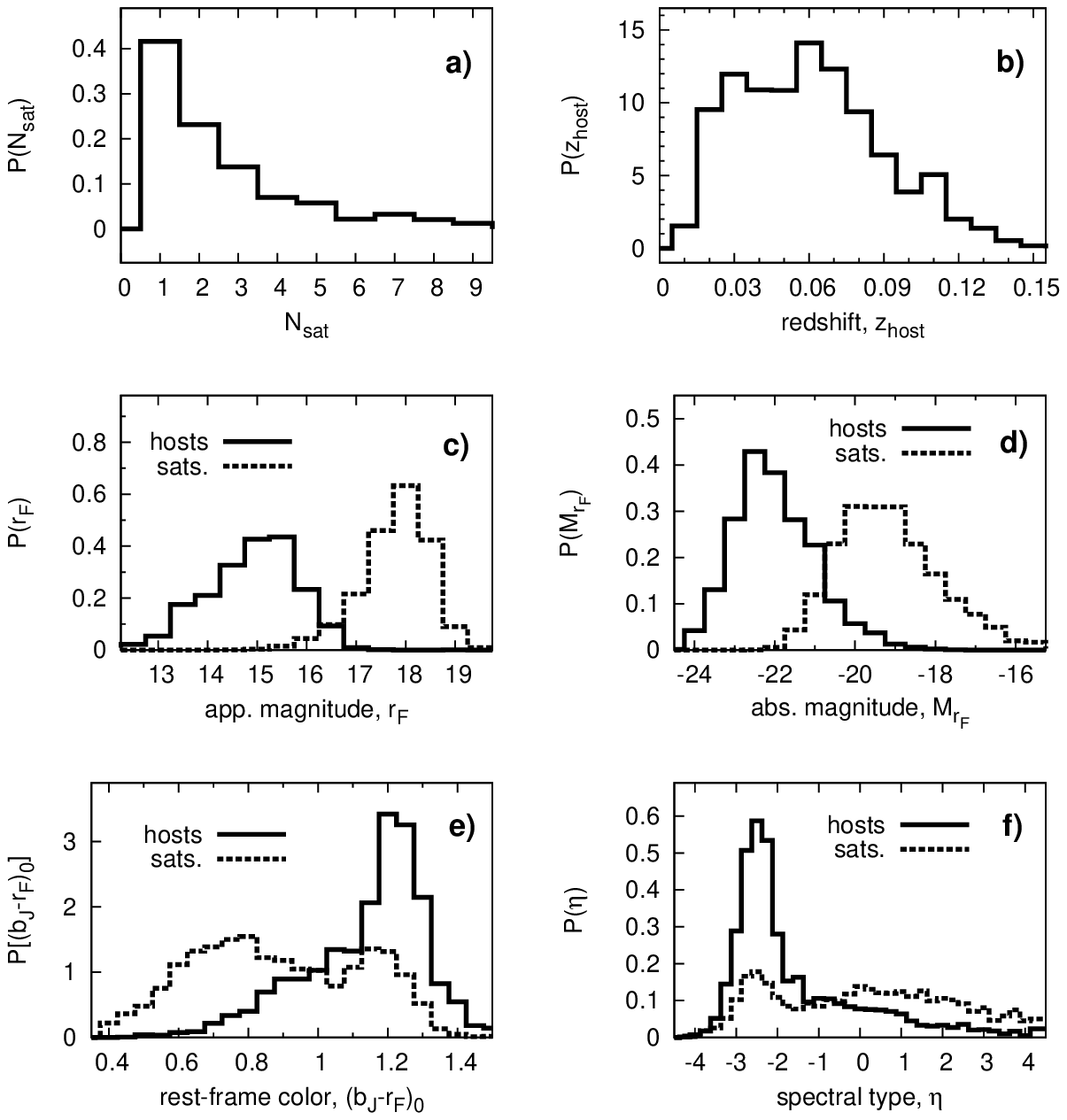} } }%
\vskip -0.0cm%
\caption{Statistical properties of Sample~2: 
a) number of satellites per host,
b) host redshifts, c) $r_F$ apparent magnitudes
of hosts (solid line) and satellites (dotted line), 
d) $r_F$ absolute magnitudes
of hosts (solid line) and satellites (dotted line), 
e) rest-frame colors of
hosts (solid line) and satellites (dotted line), 
and f) spectral types of hosts (solid line) and satellites (dotted line).
Here the magnitudes have been obtained from the SuperCOSMOS scans.
}
\label{fig2}
\end{figure}

\begin{figure}
\centerline{\scalebox{0.90}{\includegraphics{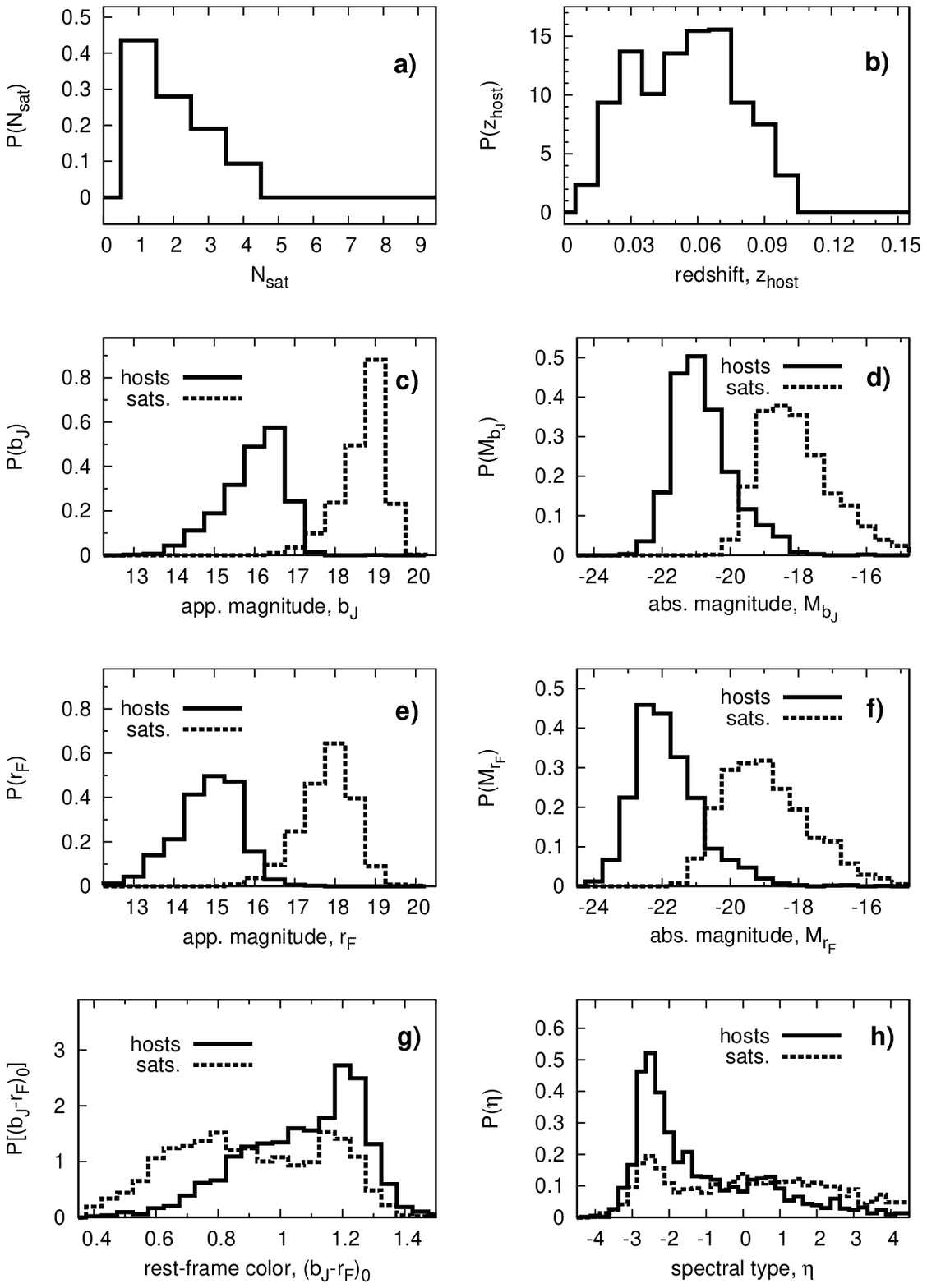} } }%
\vskip -0.0cm%
\caption{Statistical properties of Sample~1 for comparison to
Sample~2: 
a) number of satellites per host,
b) host redshifts, c) $b_J$ apparent magnitudes
of hosts (solid line) and satellites (dotted line),
d) $b_J$ absolute magnitudes
of hosts (solid line) and satellites (dotted line), 
e) $r_F$ apparent magnitudes of hosts (solid line) and
satellites (dotted line),
f) $r_F$ absolute magnitudes of hosts (solid line) and
satellites (dotted line),
g) rest-frame colors of
hosts (solid line) and satellites (dotted line), 
and h) spectral types of hosts (solid line) and satellites (dotted line).
Here the magnitudes have been obtained from the SuperCOSMOS scans.
}
\label{fig3}
\end{figure}

Probability distributions for the locations of all of the satellites in our
second sample are shown in the bottom panels of Figure~1.  The differential probability distribution,
$P(\phi)$, is shown in Figure~1c), along with the mean
satellite location and the confidence
level at which the $\chi^2$ test rejects a uniform distribution for the satellites.
The cumulative probability distribution, 
$P(\phi \le \phi_{\rm max})$, is shown in Figure~1d), along with
the median satellite location and
the confidence level at which the KS test rejects a uniform distribution for the
satellites.  From Figure~1c) and 1d), then, the 2dFGRS satellites 
in our second sample are distributed
anisotropically about their hosts, and the sense of the anisotropy is the same 
as the anisotropy of the SDSS satellites: when averaged over the entire sample,
the satellites are located preferentially close to the major axes of their hosts.
The significance of our detection of the anisotropy is, however,
less for the locations of the
2dFGRS satellites ($\chi^2$ and KS rejection confidence levels of 99.7\% 
and 99.6\%, respectively)
than it was for the locations of the SDSS satellites in our previous
study ($\chi^2$ and KS rejection 
confidence levels $> 99.99$\%; see
[11]).
This is likely due to a combination of effects.
First,  the host-satellite sample in [11] is $\sim 50$\% larger
than the one we have used here (4,487 SDSS hosts and 7,399 SDSS satellites),
which simply results in better statistics.  Second, although the SDSS and 2dFGRS
hosts have very similar redshift distributions, the images of the SDSS galaxies are
somewhat
better resolved than the images of the 2dFGRS galaxies (pixel size of 0.40~arcsec
in the SDSS vs.\ pixel size of 0.67~arcsec for the SuperCOSMOS scans).  This
could lead to more accurate centroids for the SDSS galaxies,
as well as more accurate position angles for the SDSS hosts.
In addition, the rms velocity error
in the 2dFGRS is $\sim 85$~\kms (e.g., [17]), which is significantly greater than the 
$\sim 30$~\kms~rms velocity error in the SDSS (e.g., [31]).  
As a result, it would not 
be surprising if the 2dFGRS sample contains a larger fraction of interlopers
than does the SDSS sample.  The effect of interlopers is to reduce the observed
anisotropy in the satellite locations (e.g., [11]). Hence, 
due to the smaller size of the 2dFGRS sample,
the greater interloper contamination of the 2dFGRS sample, and the more accurate
photometric parameters of the SDSS,
we would naturally expect
to find somewhat less anisotropy in the locations of the 2dFGRS satellites than in the
locations of the SDSS satellites.

Next, using our second 2dFGRS host-satellite sample we again investigate whether dividing
the sample into host-satellite pairs with low relative velocities ($| \delta v | <
160$~\kms; 1,988 hosts and 2,633 satellites) 
and high relative velocities ($| \delta v | > 160$~\kms; 1,512 hosts and 2,097 satellites) 
affects our ability to detect the anisotropy in the satellite locations.  We summarize
our results in Table~1, lines 10--12, from which it is clear that the anisotropy 
in the satellite locations
is detected for the host-satellite pairs with low relative velocities (although,
due to the smaller number of satellites, the  
significance is lower than it is for the full sample).
In the case of the host-satellite
pairs with high relative velocities, the satellite locations are consistent with an 
isotropic distribution (KS rejection confidence level of
87\%).  However, it is important to note that this alone does not
constitute proof that the locations of the satellites in the high relative velocity sample
are intrinsically different from the locations of the satellites in the low relative
velocity sample (e.g., as might be expected if the high relative velocity sample 
contained a much larger fraction of interlopers than
the low relative velocity sample).

In order to determine whether the satellite locations in the high relative velocity
sample are truly different from those in the low relative velocity sample, we again compute
a two-sample KS test.  When we compare $P(\phi \le \phi_{\rm max})$ for the high
relative velocity sample to $P(\phi \le \phi_{\rm max})$ for the low relative velocity
sample, we find that the two distributions are statistically indistinguishable; the 
two-sample KS test rejects the null hypothesis that the two distributions are drawn
from the same distribution at a 
confidence level of 54\%.  Therefore, it is not correct to conclude
that dividing our second sample by relative velocity yields one set of satellites that
are distributed anisotropically about their hosts (i.e., the low relative velocity sample)
and another set of satellites that are distributed isotropically about their hosts
(i.e., the high relative velocity sample).  

At least for the rather small host-satellite sample that can be obtained from the 
2dFGRS, it does not appear that dividing the sample by host-satellite relative
velocity yields a substantial reduction in the effects of interloper contamination
on the observed locations of the satellite galaxies.  In other words,
since $P(\phi \le \phi_{\rm max})$ for the high relative velocity sample is
consistent with being drawn from the same
distribution as $P(\phi \le \phi_{\rm max})$ for the low relative velocity sample,
there is no statistically significant evidence that the satellites in the high
relative velocity sample are distributed much more uniformly around their hosts than
are the satellites in the low relative velocity sample.
Both theoretically (e.g., [11], [22], [23]) and observationally, then, 
dividing the sample by
host-satellite relative velocity does not obviously provide a significant reduction of
the effects of interlopers on the observed locations of satellite galaxies.

If, however, we consider the star formation rates
of the satellites, it could in principle be possible to identify a sample of satellites that
contains both the smallest level of interloper contamination 
and the greatest degree of intrinsic anisotropy in the locations of
the genuine satellites. 
From the theoretical work by 
[11], we know that the selection criteria
that we have adopted here yield host galaxies that reside at the dynamical 
centers of large dark matter halos.  The satellites are non-central galaxies
(i.e., ``sub-structure'')
that orbit within their hosts' large dark matter halos.  Prior to being
accreted into the dark matter halo of its host galaxy, a satellite galaxy
would have grown and evolved within its own dark matter halo.  After accretion,
the satellite would have ceased growing in mass, and may have even lost mass
(e.g., by tidal stripping when passing near the center of its hosts' halo, or
by interactions with other subhalos).  Star formation within the satellite 
would have been severely quenched after accretion by the
host galaxy because the satellite loses most of its cold gas reservoir
to the warmer, larger halo of its host.  The higher the redshift at which
a satellite was accreted, then, the lower should be its star formation rate 
at the present day, and the more likely its orbit will reflect the (non-spherical)
gravitational potential of its host's dark matter halo.

From [22], we know that by the present day (i.e., $z = 0$), only
$\sim 40$\% of the genuine satellite galaxies in the Millennium
Run Simulation (i.e., [1]) that have 
blue SDSS colors, $(g-r)_0 < 0.7$, have completed at least one 
orbit of their host galaxy.  In contrast,
$\sim 86$\% of the genuine satellite galaxies with 
red SDSS colors, $(g-r)_0 \ge 0.7$, have completed one or more orbits
of their host galaxy by the present day.  In addition,  [11]
found that when our selection criteria above were applied
to the Millennium Run Simulation, only 42\% of the objects with blue SDSS colors
that were selected as satellites were, in fact, genuine satellites.
However, [11] also found that 81\%
of the objects with red SDSS colors that were selected as satellites 
were actually genuine satellites.  All in all, then,
we would expect that an observational sample of satellite galaxies with
low star formation rates and red SDSS colors
should suffer the least amount of interloper contamination,
while also exhibiting the greatest amount of intrinsic anisotropy in their locations 
relative to their hosts (i.e., since they are relatively
``old'' satellites that have been within their hosts' halos for a considerable
length of time).

Due to the very small overlap of the 2dFGRS and the SDSS, SDSS colors
are not available for more than a few objects in our sample.  
However, using the parameter
$\eta$ we can investigate the effects of star formation rate on the 
observed locations of the 2dFGRS satellites.  If we classify the satellites
with $\eta > -1.4$ as ``star-forming'' (3201 satellites)
and $\eta \le -1.4$ as ``passive'' (1131 satellites),
we then find that the cumulative probability distribution for the locations
of the passive satellites is inconsistent with an isotropic distribution
(KS rejection confidence level of 99\%), while the cumulative probability
distribution for the locations of the star-forming satellites is
consistent with an isotropic distribution (KS rejection confidence level
of 89.4\%).  In the case of the passive satellites, $\left< \phi \right> =
43.0^\circ \pm 0.8^\circ$ and $\phi_{\rm med} = 42.^\circ3^{+1.2}_{-1.1}$, while
for the star-forming satellites $\left< \phi \right> = 44.2^\circ \pm 0.5^\circ$
and $\phi_{\rm med} = 43.5^\circ \pm 0.7^\circ $.
As with the above results for the locations of satellites 
with high and low velocities relative to their hosts, however, this 
result should
not be interpreted as evidence that the star-forming satellites are
distributed isotropically about their hosts, while the passive satellites
are distributed anisotropically about their hosts.
Rather, a two-sample KS test finds that $P(\phi \le \phi_{\rm max})$ for
the star-forming satellites is statistically indistinguishable from
$P(\phi \le \phi_{\rm max})$ for the passive satellites (KS rejection 
confidence level 88.8\%).

Finally, it is worth noting that, unlike our first sample,  in our
second sample we find a
statistically-significant detection of anisotropic satellite locations
when we use the locations of
all of the satellites in the analysis; i.e., in our 
first sample, the satellite locations are consistent with
an isotropic distribution.  Given that our 
second sample is almost twice as large as our first sample, it is 
tempting 
to attribute the difference in the results from 
the two samples solely to improved statistics.  However, the
increase in the sample size does not appear to be the primary
cause of the increased signal-to-noise.
Instead, the selection of the hosts and 
satellites specifically using the SuperCOSMOS photometry seems to be
the source of the improved signal-to-noise in our second sample.

If we simply restrict the analysis of our second sample to only those hosts
that have $z \le 0.1$, K-corrected SuperCOSMOS absolute magnitudes
$B_J < -18$, ellipticities $\epsilon > 0.1$ as 
measured from the $r_F$ SuperCOSMOS
photometry, and fewer than 5 satellites (i.e., to effectively match
the selection criteria used to obtain our first sample),  our second sample is 
substantially reduced in size: 2,089 hosts and 3,056 satellites.  This
restricted version of our second sample is only $\sim 20\%$ larger than
our first sample, which was selected using the APM scans of 
the $b_J$ plates.  This smaller, restricted $r_F$ sample is 
substantially different
from our $b_J$ sample in Section~3 in that it includes only 1272 of
the 1725 hosts in the $b_J$ sample and only 1835 of the 2594 satellites
in the $b_J$ sample.   Therefore, $\sim 40$\% of the hosts and
satellites in the restricted $r_F$ sample are not present in the $b_J$-selected
sample from Section~3, and $\sim 25$\% of the hosts and satellites
in the $b_J$-selected sample are not present in the restricted $r_F$ sample.

When averaged over all satellites in the restricted version of
our second sample, the 
locations of the satellites are still inconsistent with an isotropic
distribution (KS rejection confidence level of 99.9\%, and $\chi^2$ rejection
confidence level of 99.6\%).  Therefore, using the 
the SuperCOSMOS scans of the $r_F$ plates allows a detection of the anisotropic
distribution of satellite galaxies that was not possible with the original
APM scans of the $b_J$ plates.

\section{Summary}

We have computed the locations of satellite galaxies in the 2dFGRS using
two sets of selection criteria, and we have investigated whether dividing the
sample by host-satellite relative velocity provides a significant reduction
of the effects of interlopers on the 
observed locations of the satellites.  When we adopt the selection criteria
used by [7] in their study of the locations of 2dFGRS satellites, we
find no statistically significant evidence that the satellites are
distributed anisotropically about their hosts.  This result is independent
of the photometric catalogs that we use (APM scans of the $b_J$ plates,
and SuperCOSMOS scans of the $b_J$ and $r_F$ plates), as well as the
velocities of the satellites relative to their hosts.  Our result is
in contrast to the original
study of [7], who found that the 2dFGRS satellites in the low 
relative velocity sample are distributed anisotropically around their hosts.
The cause of this discrepancy is not clear, but it may lie in the fact that
our samples are not truly identical, or perhaps in differences 
in the way that
the satellite locations were calculated in our independent analyzes.

We obtain a second host-satellite sample by applying
a set of selection criteria that are based upon the criteria we used
in a previous study of the locations of satellite galaxies in the SDSS.
Further, our second sample is obtained using the $r_F$ SuperCOSMOS photometry
instead of the $b_J$ APM photometry.
Using our second sample, we find that the satellites are
anisotropically distributed about their hosts at a statistically-significant
level (KS rejection confidence level of 99.6\%).  The sense of the anisotropy
is in agreement with previous studies; when averaged over the entire
population, the satellites have a preference to be found near the 
major axes of their host galaxies.  When we divide our second sample
into host-satellite pairs with low relative velocities ($| \delta v | < 160$~\kms)
and host-satellite pairs with high relative velocities ($| \delta v | > 160$~\kms),
we find that the satellites with low relative velocities are
anisotropically distributed about their hosts at a statistically-significant
level, while an isotropic distribution cannot be ruled out for
the locations of the satellites with high relative velocities.  However, this
result should not be interpreted as evidence that the distribution of the 
satellites in the low relative velocity sample is intrinsically different from
that of the satellites in the high relative velocity sample.  When we compare
the distributions of the satellites in the low and high relative velocity 
samples, we find that they are statistically indistinguishable.  As a result,
it is not clear that dividing the sample by host-satellite relative velocity
is a direct means of eliminating the effects of interlopers on the
observed locations of satellite galaxies. 

Although the selection criteria that we use to obtain our second sample
results in a sample that is nearly twice as large as our first sample, 
the increase in the sample size is not the primary reason that the anisotropy
in the satellite locations can be detected in the second sample, but not
the first.  Instead, it is the improved photometry from the SuperCOSMOS
scans of the $r_F$ plates 
that leads to the increased signal-to-noise.  If we restrict
the analysis of our second sample to a set of host-satellite systems
whose properties
match those of our first sample, the second sample is only $\sim 20$\% 
larger than the first sample, yet the anisotropy of the satellite locations
is detected at a highly-significant level (KS rejection confidence level
of 99.9\%).

Finally, in an attempt to isolate a population of satellites that are
likely to have the least interloper contamination, as well as the 
greatest degree of anisotropy in the locations of the genuine satellites,
we investigated the effects of star formation rate on the locations of the
2dFGRS satellites.  In our second sample, we find that passive satellites
(which constitute only 26\% of the satellites with well-defined spectral
parameters) are distributed anisotropically around their hosts with high
statistical significance (KS rejection confidence level of 99\%).  An isotropic
distribution cannot be ruled out for the locations of the star-forming
satellites.  However, as with our result for dividing the sample by
relative velocity, this should not be interpreted as evidence that
the locations of the star-forming satellites around their hosts are
intrinsically different from the locations of the passive satellites.  Rather,
we find that the two distributions of satellite locations are statistically
indistinguishable in our 2dFGRS sample.  Although the star formation
rates are quantified differently in the SDSS than they
are in the 2dFGRS (i.e., star formation is
quantified by SSFR, not $\eta$, in the SDSS), this
last result is in reasonable agreement with our previous results for 
satellite galaxies in the SDSS.  
That is, the SDSS satellites with the 
lowest SSFR show a pronounced tendency to be located near the major 
axes of their hosts, and the SDSS satellites with the highest SSFR show
little anisotropy in their locations.  However, the mean satellite locations,
$\left< \phi \right>$, for the SDSS satellites with the highest SSFR
 and the lowest
SSFR agree with each other at the $2\sigma$ level (see Figure~11c) in
[11]).  Therefore, it is not clear that dividing the sample by
the star formation rates of the satellites is sufficient to 
largely eliminate
the effects of interlopers on the observed locations of satellite 
galaxies.

\section*{Acknowledgments}

We are pleased to thank all of those who
were involved with the 2dFGRS.  
This work was supported by the National Science Foundation 
under NSF contract AST-0708468.

\end{document}